\documentclass[12pt]{article}

\usepackage{amssymb}
\usepackage{amsmath}
\usepackage{array} 
\usepackage{epsfig}
\usepackage{graphics}

\begin{document}

\begin{center}

 {\Large \bf
\vskip 7cm
\mbox{Towards Extraction of $\pi^+\; p$ and $\pi^+\pi^+$ cross-sections}\\
\mbox{from Charge Exchange Processes at the LHC.}\\
}

\vskip 1cm

\mbox{R.A.~Ryutin, V.A.~Petrov, A.E.~Sobol}\\
\vskip 0.5cm
\mbox{{\small Institute for High Energy Physics}}
\mbox{{\small{\it 142 281} Protvino, Russia}}\\
\vskip 0.5cm

 \vskip 1.75cm
{\bf
\mbox{Abstract}}
  \vskip 0.3cm

\newlength{\qqq}
\settowidth{\qqq}{In the framework of the operator product  expansion, the quark mass dependence of}
\hfill
\noindent
\begin{minipage}{\qqq}
We study the possibilities to analyse the data on leading neutrons production at first LHC runs. These data could be used to extract from it $\pi^+ p$ and $\pi^+\pi^+$ cross-sections. In this note we estimate relative contributions of $\pi$, $\rho$ and $a_2$ reggeons to charge exchanges and discuss related problems of measurements.

\end{minipage}
\end{center}


\begin{center}
\vskip 0.5cm
{\bf
\mbox{Keywords}}
\vskip 0.3cm

\settowidth{\qqq}{In the framework of the operator product  expansion, the quark mass dependence of}
\hfill
\noindent
\begin{minipage}{\qqq}
Leading Neutron Spectra -- Elastic and total cross-section -- Absorption -- Regge-eikonal model - pion-proton and pion-pion collider
\end{minipage}

\end{center}
\setcounter{page}{1}
\newpage

\section{Introduction}
In recent papers~\cite{ourneutrontot},\cite{ourneutronel} we pushed forward (and discussed)
the idea of using the Zero
Degree Calorimeters~\cite{ZDC},  ZDCs, designed for different uses at 
several of the LHC collaborations, to extract the total and elastic 
cross-sections of the $\pi^+ p$ and $\pi^+\pi^+$ scattering processes.  Actually, this 
could allow the use of the LHC as a $\pi p$ and $\pi\pi$ collider at effective c.m.s. energies 
about 1-5 TeV. For further motivation and technical details we refer the reader to 
Refs.~\cite{ourneutrontot},\cite{ourneutronel}.

In this paper we concentrate on quite a serious problem of the $\rho$- and $a_2$-exchanges in the
processes $p+p\to n+X$ and $p+p\to n+X+n$ which compete with the $\pi^+$-exchange and are to be considered in detail.

\section{The basic model of charge exchange processes and extraction of $\pi^+$ $p$ and $\pi^+$ $\pi^+$ cross-sections.}

\begin{figure}[b!]
  \hskip 4cm
  \vbox to 5.5cm
 {\hbox to 8cm{\epsfxsize=8cm\epsfysize=5.5cm\epsffile{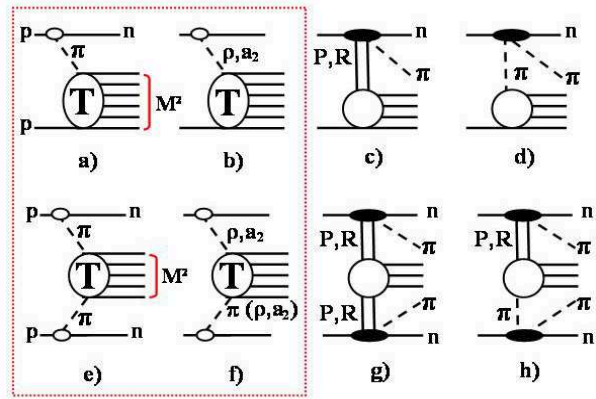}}}
 \caption{\small\it\label{fig:diags_all} Signal and background processes: a) S$\pi$E signal; b) SRE background; c)d) Double Dissociative (DD) background; e) D$\pi$E signal; f) DRE background (contributions from $\pi\;\rho$ and $\pi\; a_2$ collisions dominate); g)h) Central Diffractive (CD) background.}
 \end{figure}

 We consider processes presented in Fig.~\ref{fig:diags_all}. Signal processes 
 of Single (S$\pi$E) and double (D$\pi$E) pion exchanges are depicted in
 Fig.~\ref{fig:diags_all} a), e). In the previous
 articles~\cite{ourneutrontot},\cite{ourneutronel} we estimated contributions
 to the background of reactions depicted in 
 Fig.~\ref{fig:diags_all}c)d)g)h) and also 
 minimum bias (MB) and single dissociation (SD) with forward neutrons 
 production. In the present work we give calculations of events from
 Fig.~\ref{fig:diags_all} b), f), which
 are called single (SRE) and double (DRE) reggeon exchanges. In the DRE 
 contributions of $\pi\;\rho$ and $\pi\; a_2$ collisions dominate over $\rho\;\rho$, $\rho\; a_2$ and $a_2\; a_2$ processes.

\begin{center}
\begin{figure}[t!]
\hskip  3cm \vbox to 5.5cm
{\hbox to 9cm{\epsfxsize=9cm\epsfysize=5.5cm\epsffile{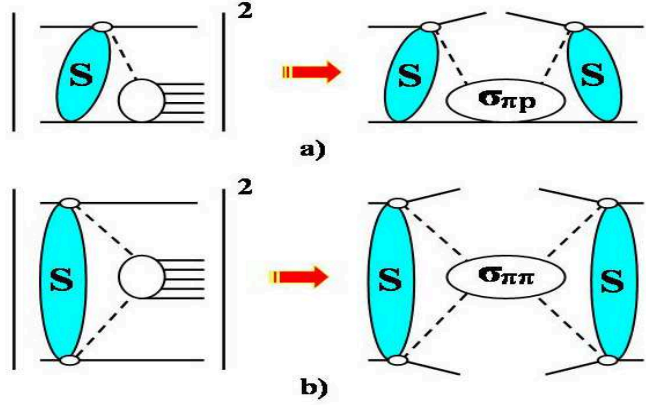}}}
\caption{\small\it\label{fig:diags_cs} Amplitudes squared and cross-sections of the processes: a) $p+p\to n+X$ (S$\pi$E), b) $p+p\to n+X+n$ (D$\pi$E). $S$ represents soft rescattering corrections.}
\end{figure}
\end{center}

\begin{center}
\vskip -2.35cm
\end{center} 
 Details of calculations can be found in~\cite{ourneutrontot},\cite{ourneutronel}. Here we show only basic issues. As an approximation for $\pi$ exchanges we use the formulas shown graphically in Fig.~\ref{fig:diags_cs}. If we take into account absorptive corrections, which were calculated in the Regge-eikonal model~\cite{3Pomerons}, these formulas can be rewritten as 
\begin{equation}
 \label{csSpiE}\frac{d\sigma_{{\rm S}\pi {\rm E}}}{d\xi dt}=F_0(\xi, t)S(s/s_0, \xi, t)\;\sigma_{\pi^+ p}(\xi s) ,
\end{equation}
\begin{equation}
 \label{csDpiE} \frac{d\sigma_{{\rm D}\pi {\rm E}}}{d\xi_1d\xi_2dt_1dt_2}=F_0(\xi_1, t_1)F_0(\xi_2,t_2)S_2(s/s_0, \{\xi_i\},\{t_i\})\;\sigma_{\pi^+\pi^+}(\xi_1\xi_2 s),
\end{equation}
\begin{equation}
\label{F0form} F_0(\xi,t)=\frac{G_{\pi^+pn}^2}{16\pi^2}\frac{-t}{(t-m_{\pi}^2)^2} {\rm e}^{2bt} \xi^{1-2\alpha_{\pi}(t)},
\end{equation}
where the pion trajectory is $\alpha_{\pi}(t)=\alpha^{\prime}_{\pi}(t-m_{\pi}^2)$. The slope $\alpha^{\prime}_{\pi}\simeq 0.9$~GeV$^{-2}$, $\xi=1-x_L$, where $x_L$ is the fraction of the initial proton's longitudinal momentum carried by the neutron, and $G_{\pi^0pp}^2/(4\pi)=G_{\pi^+pn}^2/(8\pi)=13.75$~\cite{constG}. From recent
data~\cite{HERA2},\cite{KMRn1c16}, we expect $b\simeq 0.3\; {\rm GeV}^{-2}$. 
We are interested in the kinematical range $0.01$~GeV$^2<|t_i|<0.5$~GeV$^2$, $\xi_i<0.4$, where formulaes~(\ref{csSpiE}),(\ref{csDpiE}) dominate according to~\cite{KMRn1c13} and~\cite{KMRn1c14}.

\begin{center}
\begin{figure}[hb!]
\vbox to 5cm
{\hbox to 15cm{\epsfxsize=15cm\epsfysize=5cm\epsffile{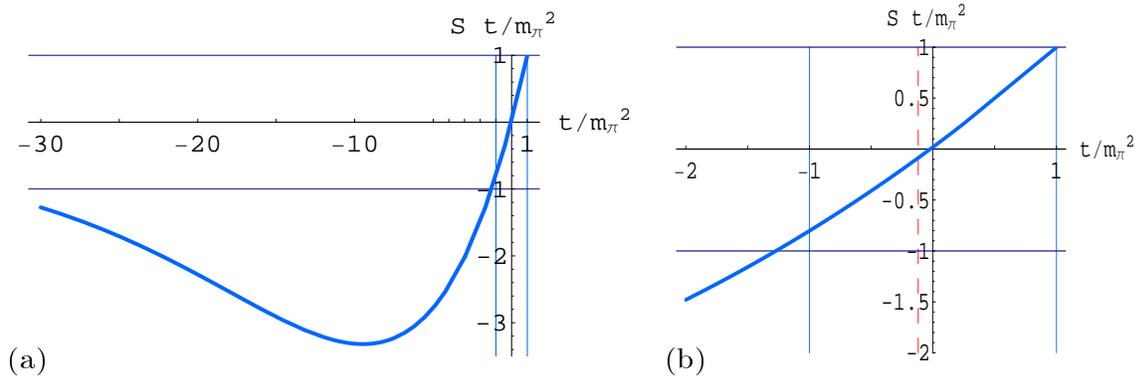}}}
\vskip 0.8cm
\caption{\small\it\label{fig:St0} Function $S(\xi,t)\;t/m_{\pi}^2$ versus $t/m_{\pi}^2$ at fixed $\xi=0.05$. The boundary of the physical region $t_0=-m_p^2\xi^2/(1-\xi)$ is represented by 
vertical dashed line in b).}
\end{figure}
\end{center}

Rescattering corrections $S$ and $S_2$ are calculated in~\cite{ourneutrontot},\cite{ourneutronel}. Behaviour of $S\;t/m_{\pi}^2$ is shown in the Fig.~\ref{fig:St0}. It is clear from the figure 
that $|S|\sim 1$ at $|t|\sim m_{\pi}^2$ (the situation is similar for $S_2$), which is an argument for the possible
model-independent extraction of $\pi p$ and $\pi\pi$ cross-sections by the use of~(\ref{csSpiE}) and~(\ref{csDpiE})~\cite{ourneutronel}.
 
\begin{center}
\begin{figure}[hb!]
\hskip 1cm\vbox to 5cm
{\hbox to 14cm{\epsfxsize=14cm\epsfysize=5cm\epsffile{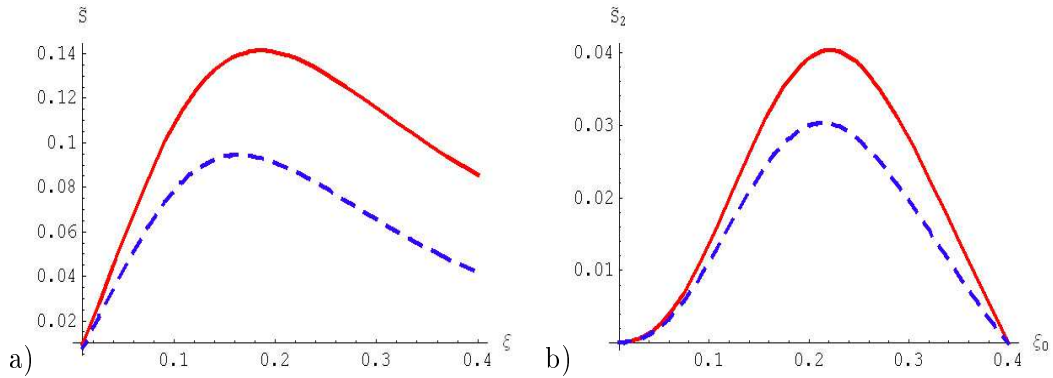}}}
\vskip 0.8cm
\caption{\small\it\label{fig:survint} Rescattering corrections integrated with formfactors for $\sqrt{s}=0.9$~TeV (solid) and $\sqrt{s}=7$~TeV (dashed): a) $\tilde{S}(s,\;\xi)$; b) $\tilde{S}_2(s,\;\xi_0)$.}
\end{figure}
\end{center}

The present design of detectors does not allow $t$ measuremets, it gives only restrictions $|t|<\sim1\;{\rm GeV}^2$ at $7$~TeV ($|t|<0.3\;{\rm GeV}^2$ at $0.9$~TeV). If to 
assume a weak enough $t$-dependence of $\pi p$ and $\pi\pi$ cross-sections, i.e.
\begin{equation}
\label{virtpireal}
\sigma_{\pi^+_{virt}p}(s; \{ m_p^2, t \})\simeq \sigma_{\pi^+ p}(s; \{ m_p^2, m_{\pi}^2 \}),\; 
\sigma_{\pi^+_{virt}\pi^+_{virt}}(s; \{ t_1,t_2 \})|\simeq \sigma_{\pi^+ \pi^+}(s; \{ m_{\pi}^2, m_{\pi}^2 \}),
\end{equation}
then we could hope to extract these cross-sections (though, with big errors) by the following procedure:
\begin{eqnarray}
  \label{eq:pipextractINT}
 \tilde{S}(\xi)&\!\!\!\!=&\!\!\!\!\int\limits_{t_{min}}^{t_{max}}dt \;S(s/s_0,\xi,t)F_0(\xi,t),\; \nonumber\\
  \sigma_{\pi^+ p}(\xi s)&\!\!\!\!=&\!\!\!\!\frac{\frac{d\sigma_{{\rm S}\pi {\rm E}}}{d\xi}}{\tilde{S}(s,\xi)},\; \xi\simeq\frac{M_{\pi p}^2}{s},\\
  \tilde{S}_2(\xi_0)&\!\!\!\!=&\!\!\!\!\int\limits_{t_{min}}^{t_{max}}dt_1 dt_2\int\limits_{-y_0}^{y_0}dy\; S_2(s/s_0, \{\xi_0 {\rm e}^{\pm y}\}, \{t_i\})F_0(\xi_0 {\rm e}^y,t_1)F_0(\xi_0{\rm e}^{-y},t_2),\nonumber\\
 \label{eq:pipiextractINT}
\sigma_{\pi^+ \pi^+}(\xi_0^2 s)&\!\!\!\!=&\!\!\!\!\frac{\frac{d\sigma_{{\rm D}\pi {\rm E}}}{d\xi_0}}{\tilde{S}_2(s,\xi_0)},\; \xi_0=\frac{M_{\pi\pi}}{\sqrt{s}},\; y_0=\ln\frac{\xi_{\rm max}\sqrt{s}}{M_{\pi\pi}}.
\end{eqnarray}
Functions $\tilde{S}_2(s,\;\xi_0)$ and $\tilde{S}(s,\;\xi)$ are depicted in Fig.~\ref{fig:survint}. To  suppress theoretical errors of $\tilde{S}$ and $\tilde{S}_2$  we have to measure total and elastic $pp$ rates at energies greater than $2$~TeV, since all the models for absorptive corrections are normalized to $pp$ cross-sections. At present we can estimate the theoretical error to be less than 20\% at $\sim 10$~TeV for this method from predicted values of total $pp$ cross-sections in the most popular models~\cite{ourneutronel}.
  
 For $\rho$ and $a_2$ contributions we can write formulaes similar to~(\ref{csSpiE}), (\ref{csDpiE}):
\begin{eqnarray}
\label{csSRE}\frac{d\sigma_{\rm SRE}}{d\xi dt}&=&F_R(\xi, t)S_R(s/s_0, \xi, t)\;\sigma_{R^+ p}(\xi s) ,\\
 \label{csDRpiE} \frac{d\sigma_{{\rm DR}\pi {\rm E}}}{d\xi_1d\xi_2dt_1dt_2}&=&F_{R\pi}(\xi_1, \xi_2,t_1, t_2)S_{R,2}(s/s_0, \{\xi_i\},\{t_i\})\;\sigma_{R^+\pi^+}(\xi_1\xi_2 s),
\end{eqnarray}
\begin{eqnarray}
 \label{FRform} F_R(\xi,t)&=&\frac{|\eta_R|^2\tilde{G}_{R^+pn}^2}{16\pi^2}{\rm e}^{2b_Rt}\xi^{1-2\alpha_R(t)}\left(1+\kappa_R^2\frac{\vec{q}^{\;2}}{4m_p^2}\right),\\
 F_{R\pi}(\{\xi_i\},\{t_i\})&=&F_0(1)F_R(2)+F_0(2)F_R(1)+2\sqrt{\frac{F_0(1)F_0(2)F_R(1)F_R(2)}{t_1t_2(1-\xi_1)(1-\xi_2)}}\nonumber\\
 &\times& \frac{\left(m_p\xi_1+\vec{q}_{1}^{\;2}\frac{\kappa_R}{2m_p}\right)\left(m_p\xi_2+\vec{q}_{2}^{\;2}\frac{\kappa_R}{2m_p}\right)}{\left(1+\vec{q}_{1}^{\;2}\frac{\kappa_R^2}{4m_p^2}\right)\left(1+\vec{q}_{2}^{\;2}\frac{\kappa_R^2}{4m_p^2}\right)},\\
 F_{0,R}(i)&\!\!\!\!=&\!\!\!\! F_{0,R}(\xi_i,t_i),\; \vec{q}_i^{\;2}\simeq-t_i(1-\xi_i)-m_p^2\xi_i^2.
\end{eqnarray}
Here $\kappa_R=8$ is the ratio of spin-flip to nonflip amplitude, $\alpha_R(t)\simeq 0.5+0.9t$ and parameters for $\rho$, $a_2$ mesons are~\cite{rhoa2pars}
\begin{eqnarray}
&& \eta_{\rho}=-\imath+1,\; \eta_{a_2}=\imath+1,\\
&& b_{\rho}=2\;{\rm GeV}^{-2},\; b_{a_2}=1\;{\rm GeV}^{-2},\\
&& \frac{\tilde{G}_{\rho^+pn}^2}{8\pi}=0.18\;{\rm GeV}^{-2},\; \frac{\tilde{G}_{{a_2}^+pn}^2}{8\pi}=0.405\;{\rm GeV}^{-2}.
\end{eqnarray}
Rescattering corrections $S_R$ and $S_{R,2}$ are calculated by the method used in~\cite{ourneutrontot},\cite{ourneutronel}.
Basic assumptions in our calculations are: 
\begin{itemize}
\item $\rho\;\rho$, $\rho\; a_2$ and $a_2\; a_2$ contributions are small; 
\item interference terms of the type $T^*_{{\rm S}\pi{\rm E}}T_{{\rm SRE}}$, $T^*_{{\rm DR}\pi{\rm E}}T_{{\rm DR}^{\prime}\pi{\rm E}}$ are small~\cite{KMRn1c16}, ${\rm R, R}^{\prime}=\pi,\;\rho,\;a_2$, $R\neq R^{\prime}$, where $T$ are amplitudes of the corresponding processes;
\item approximate relations $\sigma_{R^+ p}\simeq\sigma_{\pi^+ p}$, $\sigma_{R^+\pi^+}\simeq\sigma_{\pi^+\pi^+}$~\cite{KMRn1c16}.
\end{itemize}

\section{Relative contributions of $\pi$, $\rho$ and $a_2$ exchanges to charge exchanges}

 Let us consider meson exchange contributions as a source of 
additional backgrounds for S$\pi$E and D$\pi$E. In Figs.~\ref{fig:all900} and~\ref{fig:all7000} you can see contributions of pion and reggeon (sum of $\rho$ and $a_2$) exchanges to single (CE) and double (DCE) charge exchange proceses. Here we use
the kinematical variable $r$ which is equal to the transverse distance from the beam and directly related
to the pseudorapidity $r=L/{\rm sh}(\eta)$. $L=14000\; {\rm cm}$ is the longitudinal distance from the interaction point to the 
detector. The best situation is observed at $\sqrt{s}=900\; {\rm GeV}$. Since the geometrical acceptance of the
detector is $r\le 5\;{\rm cm}$ it cuts off reggeon background almost at all for the CE (Fig.~\ref{fig:all900}b) and the significant 
part for the DCE (Fig.~\ref{fig:all900}d). At $7\; {\rm TeV}$ the situation is not so good for DCE even if we peform 
a cut $r\le 1\;{\rm cm}$ (see~Fig.~\ref{fig:all7000}d). It is difficult to separate different reggeon contributions from DCE in this case.
\begin{center}
\begin{figure}[h!]
\hskip 1cm\vbox to 8cm
{\hbox to 9cm{\epsfxsize=9cm\epsfysize=8cm\epsffile{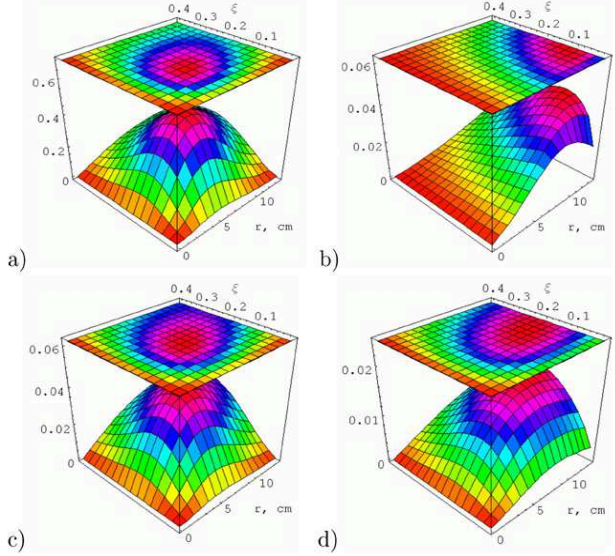}}}
\caption{\small\it\label{fig:all900} Cross-sections $\frac{d\sigma}{d\xi dr}$ in $mb\cdot cm^{-1}$ at $\sqrt{s}=0.9$~TeV for: a) S$\pi$E; b) S$\rho$E+S$a_2$E; c) D$\pi$E; d) D$\rho\pi$E+D$a_2\pi$E.}
\end{figure}
\begin{figure}[h!]
\hskip 1cm\vbox to 8cm
{\hbox to 9cm{\epsfxsize=9cm\epsfysize=8cm\epsffile{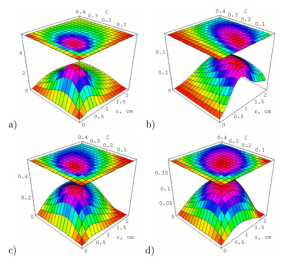}}}
\caption{\small\it\label{fig:all7000} Cross-sections $\frac{d\sigma}{d\xi dr}$ in $mb\cdot cm^{-1}$ at $\sqrt{s}=7$~TeV for: a) S$\pi$E; b) S$\rho$E+S$a_2$E; c) D$\pi$E; d) D$\rho\pi$E+D$a_2\pi$E.}
\end{figure}
\end{center}  

\begin{center}
\begin{figure}[t!]
\hskip 3cm\vbox to 4.5cm
{\hbox to 9cm{\epsfxsize=9cm\epsfysize=8cm\epsffile{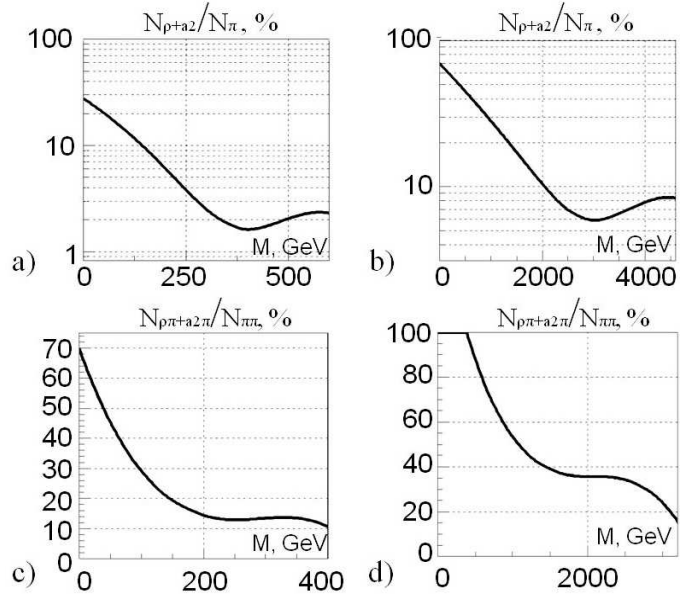}}}
\vskip 3.4cm
\caption{\small\it\label{fig:ratrtos} Ratios of reggeon exchange events to pion exchange events in the ZDC acceptance versus the invariant mass of reggeon-proton (reggeon-reggeon) systems: a) $(N_{{\rm S}\rho{\rm E}}+N_{{\rm S}a_2{\rm E}})/N_{{\rm S}\pi{\rm E}}$, $\sqrt{s}=900\;{\rm GeV}$; b) $(N_{{\rm S}\rho{\rm E}}+N_{{\rm S}a_2{\rm E}})/N_{{\rm S}\pi{\rm E}}$, $\sqrt{s}=7\;{\rm TeV}$; c) $(N_{{\rm D}\rho\pi{\rm E}}+N_{{\rm D}a_2\pi{\rm E}})/N_{{\rm D}\pi{\rm E}}$, $\sqrt{s}=900\;{\rm GeV}$; d) $(N_{{\rm D}\rho\pi{\rm E}}+N_{{\rm D}a_2\pi{\rm E}})/N_{{\rm D}\pi{\rm E}}$, $\sqrt{s}=7\;{\rm TeV}$. Results for different models are similar.}
\end{figure}
\end{center}  
\vspace*{-3cm}
\begin{table}[ht!]
\caption{\small\it\label{tab:ratios} Relative contributions of reggeons to CE and DCE in the ZDC acceptance.}
\begin{center}
\begin{tabular}{|c|c|c|c|}
\hline
 \multicolumn{2}{|c|}{$\sqrt{s},\; {\rm TeV}$}  &  0.9 & 7 \\
\hline
 \multicolumn{2}{|c|}{$(\sigma_{S\rho E}+\sigma_{Sa_2E})/\sigma_{S\pi E}$, \%}  &   10.7  &   8.2 \\
\hline
  & S$\pi$E & 27.8 & 86.6\\
ZDC acceptance, \% & S$\rho$E & 10.8 & 86.8\\
  & S$a_2$E & 6.7 & 86.7\\
\hline
\multicolumn{2}{|c|}{$\left< (N_{S\rho E}+N_{Sa_2E})/N_{S\pi E}\right>$, \%} & 3.0 & 8.2\\
\hline
\hline
 \multicolumn{2}{|c|}{$(\sigma_{D\rho\pi E}+\sigma_{Da_2\pi E})/\sigma_{D\pi\pi E}$, \%}  &   47.3  &   43.4 \\
\hline
  & D$\pi\pi$E & 4.80 & 99.6\\
ZDC acceptance, \% & D$\rho\pi$E & 0.28 & 99.8\\
  & D$a_2\pi$E & 0.65 & 99.7\\
\hline
\multicolumn{2}{|c|}{$\left< (N_{D\rho\pi E}+N_{Da_2\pi E})/N_{D\pi\pi E}\right>$, \%} & 19.3 & 43.4\\
\hline
\end{tabular}
\end{center}
\end{table}
\vspace*{-0.5cm}
Monte-carlo simulation shows relative contributions in detail (see Fig.~\ref{fig:ratrtos} and Table~\ref{tab:ratios}). For CE situation is
quiet encouraging, since reggeon background is less than 10\% for large invariant masses (or $\xi$), and for DCE it can reach 19.3 (43.4)\% at $\sqrt{s}=0.9 (7)\;{\rm TeV}$ due to similar distributions in r.
 
\section{Conclusions}

In this article we have considered the problems due to extra reggeon exchanges which arise when trying 
to extract $\pi^+\;p$ and $\pi^+\;\pi^+$ cross-sections from the data on leading neutrons at the 
LHC. After the estimation of reggeon exchange contributions to the background we can conclude 
that at present time we have some chances to extract total $\pi^+\; p$ cross-sections from the first LHC data at 900~GeV (7~TeV) 
but with rather big errors (about 20-30\%). With the data on $p\;p$ total and elastic cross-sections at 7~TeV and higher
theoretical errors can be reduced significantly.

At present our preliminary analysis shows that the 900~GeV data
are too poor to come to some valuable results. This is the reason that detectors like ZDC need modernization to improve
their performance for the reach of $\pi p$ and $\pi\pi$ collisions at the LHC.

\section*{Acknowledgements}

This work is supported by the grant RFBR-10-02-00372-a.

\end{document}